\documentclass{ws-p8-50x6-00}

\begin{document}

\title{Three-cluster nuclear molecules}

\author{D. N. Poenaru and  B. Dobrescu}

\address{Horia Hulubei National Institute of Physics and Nuclear Engineering, 
\\P.O. Box MG-6, RO-76900 Bucharest, Romania\\E-mail: poenaru@ifin.nipne.ro}

\author{W. Greiner}

\address{Institut f\"ur Theoretische Physik der Universit\"at, Postfach
111932, \\D-60054 Frankfurt am Main, Germany\\
E-mail: greiner@th.physik.uni-frankfurt.de}  

\maketitle

\abstracts{A three-center phenomenological model able to explain, at least 
from a
qualitative point of view, the difference in the observed yield of a
particle-accompanied fission and that of binary fission was developed.
It is derived from the liquid drop model under the assumption that the
aligned  configuration, with the emitted particle between the
light and heavy fragment is obtained by increasing continuously
the separation distance, while the radii of the light fragment and of the
light particle are kept constant. During the first stage of the deformation
one has a two-center evolution until the neck radius becomes equal to the 
radius of the emitted particle. Then the three center starts developing by
decreasing with the same amount the two tip distances. 
In such a way a second minimum, typical
for a cluster molecule, appears in the deformation energy.
Examples are presented for $^{240}$Pu parent nucleus emitting
$\alpha$-particles and $^{14}$C in a ternary process. }

\section{Introduction}

Fission approach\cite{p195b96b} to the cluster 
radioactivities\cite{ps84sjpn80} and $\alpha $-decay has been
systematically developed during the last two decades (see Ref. 1 
\nocite{p195b96b} and the references therein) as an alternative to the 
many-body theory.\cite{ble96mb}
One has to stress the quantum nature of these decay modes and of the 
fission process as well. The three groups of binary phenomena are taking 
place by tunneling through a potential barrier.
Fission theory has also been extended toward extremely large mass 
asymmetry\cite{mor75np} to study the evaporation of light particles 
from a hot excited compound nucleus, going over the barrier.
In a cold binary fission\cite{sig81jpl,p140bIII89b,gon91b} the 
fragments and the parent are neither
excited nor strongly deformed, hence no neutron is evaporated; the
total kinetic energy of the fragments equals the released energy.

A more complex phenomenon, the particle-accompanied fission (or ternary 
fission) was
observed both in neutron-induced and spontaneous fission. It was
discovered\cite{san46cr,mut96mb} in 1946.
Several such processes, in which the charged particle is a proton, deuteron, 
triton, $^{3-6,8}$He, $^{6-11}$Li,
$^{7-14}$Be, $^{10-17}$B, $^{13-18}$C, $^{15-20}$N,   $^{15-22}$O,
have been detected. Many other heavier isotopes of F, Ne, Na, Mg, Al, Si, P,
S, Cl, Ar, and even Ca were mentioned.\cite{gon97nc}

A very powerful technique, based on the fragment identification by using
triple $\gamma $ coincidences in the large arrays of Ge-detectors, like
GAMMASPHERE, was employed to discover new
characteristics of the fission process,\cite{ham94jpg,ter96prl}
and new decay modes\cite{ram98pr} (emission of an alpha particle and of 
$^{10}$Be, accompanying the cold fission of $^{252}$Cf,
the double fine structure, and the triple fine structure in binary and
ternary fission, respectively).

The possibility of a whole family of new decay modes, {\em the multicluster 
accompanied fission}, was recently envisaged.\cite{p215jpg99,p218pr99}
Besides the fission into two or three fragments, a heavy or 
superheavy nucleus spontaneously breaks into four, five or six
nuclei of which two are asymmetric or symmetric heavy fragments and the
others are light clusters, e.g. $\alpha$-particles, $^{10}$Be, $^{14}$C, 
$^{20}$O, or combinations of them.
Examples were presented for the two-, three- and four cluster accompanied
cold fission of $^{252}$Cf and $^{262}$Rf, in which the emitted clusters 
are:  2$\alpha$, $\alpha +
^6$He, $\alpha + ^{10}$Be, $\alpha + ^{14}$C, 3$\alpha$, $\alpha + 
^6$He~+~$^{10}$Be, 2$\alpha + ^6$He, 2$\alpha + ^8$Be, 2$\alpha + ^{14}$C, 
and 4$\alpha$.

The strong shell effect corresponding to
the doubly magic heavy fragment $^{132}$Sn was emphasized.
From the analysis of different configurations of fragments in touch,
we concluded that the most favorable mechanism of such a decay mode
should be the cluster emission from an elongated neck formed between the two
heavy fragments. The fact that the potential barrier height is lower,
suggests that in a competition between aligned and compact configurations,
the former should prevail. 

This idea is further exploited in the following for ternary fission, by 
suggesting a formation mechanism of the touching configuration, 
based on a three-center phenomenological model,
able to explain the difference in the observed yield of a
particle-accompanied fission and that of binary fission.
It is derived from the liquid drop model under the assumption that the
aligned  configuration, with the emitted particle between the
light and heavy fragment is obtained by increasing continuously
the separation distance, while the radii of the heavy fragment and of the
light particle are kept constant. During the first stage of the deformation
one has a two-center evolution until the neck radius becomes equal to the 
radius of the emitted particle. Then the three center starts developping by
decreasing with the same amount the two tip distances. 
We shall show that in such a way a second minimum, typical
for a cluster molecule, appears in the deformation energy.

\section{Shape Parametrization}

The basic condition to be fulfilled in the ternary decay process,
$^AZ \rightarrow \sum_1^3{^{A_i}Z_i}$, 
concerns the released energy ($Q$-value)
\begin{equation}
Q = M - \sum_1^3{m_i}
\end{equation}
 which should be
positive and high enough in order to  assure a relatively low potential
barrier height. The hadron numbers are conserved.  We 
took\cite{p209adndt98,p172b96a} the masses (in units of energy), entering 
in the above equation, from the compilation of
measurements.\cite{aud95np}
We make the convention $A_1 \geq A_2 \geq A_3 $.

For the first stage of the process, we adopt the shape parametrization 
of two intersected spheres with radii $R_1$
and $R_2$. By placing the origin in the center of the large sphere,
the surface equation can be written in a cylindrical system of coordinates 
as:
\begin{equation}
\rho_s^2 = \left \{ \begin{array}{lll} 
\rho_{sl}^2=&R_1^2-z^2 & \;\;\;,\;\; \mbox{$-R_1\leq z\leq z_{s1}$} \\ 
\rho_{sr}^2=&R^2_2-(z-R)^2 & \;\;\;,\;\; \mbox{$z_{s1}\leq z\leq R+R_2$} 
\end{array} \right.
\end{equation}
in which $z_{s1}$ is the position of the separation plane, 
and  $R$ is the distance between the two centers. This equation is valid as
long as $R \leq R_{ov3}$ defined below.                             
The fragment radius, $R_1$, is kept constant during the deformation, and
for a given separation distance, $R$, the radius $R_1$ is derived from 
the volume conservation and matching conditions.
\begin{figure}[ht]
\centerline{\epsfxsize=12cm\epsffile{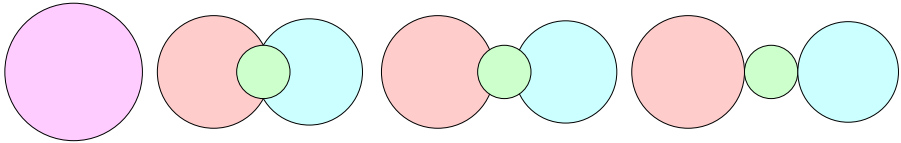}}
\caption{The assumed sequence of aligned shapes for the ternary fission
of $^{240}$Pu, leading to $^{14}$C accompanied cold fission with $^{132}$Sn
and $^{98}$Sr fragments.}
\label{Figure 1}
\end{figure}
The final fragments and the initial parent nucleus are assumed to posses 
spherical shapes with radii $R_1$, $R_2$, $R_3$, and $R_0$, where  
$R_j=1.2249A_j^{1/3}$~fm ($j=0, 1, 2, 3$). 
Within the range of $R$ from $R_i=R_0-R_1$ up to 
$R_{ov3}$ one has a configuration of two overlapping spheres. 

At $R=R_{ov3}$ (see the second position in Fig. 1) the neck radius 
$\rho_{neck1}=R_3$; $R_3$ is also kept constant. From that
moment, the third fragment comes into play and one has two necks and two
separating planes instead of one, hence:
\begin{equation}
\rho_s^2 = \left \{ \begin{array}{lll} 
\rho_{sl}^2=&R_1^2-z^2 & \;\;\;,\;\; \mbox{$-R_1\leq z\leq z_{s1}$} \\ 
\rho_{sc}^2=&R^2_3-(z-z_3)^2 & \;\;\;,\;\; \mbox{$z_{s1}\leq z\leq z_{s2}$}
\\ \rho_{sr}^2=&R^2_2-(z-R)^2 & \;\;\;,\;\; \mbox{$z_{s2}\leq z\leq R+R_2$} 
\end{array} \right.
\end{equation}
In order to arrive safely at the
final aligned configuration of fragments in touch with a corresponding
decrease of the neck radii $\rho_{neck1}$ and $\rho_{neck2}$, 
we assume a further
elongation with a corresponding decrease of the neck radii $\rho_{neck1}$
and $\rho_{neck2}$ in a particular way, 
allowing to have the same (smaller and smaller) tip
distance between the (overlapping) fragments 13 and 32 when $R$ increases
from $R_{ov3}$ to $R_{t}=R_1+R_{2f}+2R_3$.
In such a way the geometry is perfectly determined by giving one independent
shape parameter, $R$, and the mass numbers of the parent and fragment
nuclei.
\begin{figure}[ht]
\centerline{\epsfxsize=12cm\epsffile{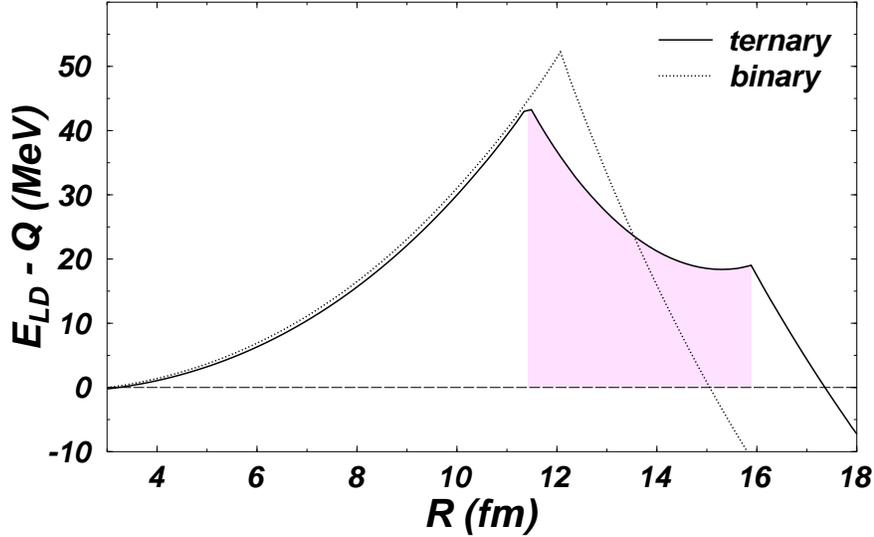}}
\caption{Deformation energy for the binary and ternary (accompanied by
$\alpha$ emission) fission of 
$^{240}$Pu. In both cases the heavy fragment is the double magic nucleus
$^{132}_{50}$Sn$_{82}$. The light fragments for binary- and ternary
processes are $^{108}$Ru and $^{104}$Mo, respectively. The region due to the
devolopment of the light particle,  from $R_{ov3}$ to $R_{t}$ is
emphasized.}
\label{Figure 2}
\end{figure}

\section{Deformation Energy}

According to the liquid-drop model (LDM),\cite{mye66np}
by requesting zero energy for a spherical shape, the
deformation energy is defined as
\begin{equation}
E_{def} = (E_s - E_s^0)+(E_C- E_C^0) = E_s^0[B_s - 1 + 2X(B_C -1)]
\end{equation}
where $E_s^0=a_s(1-\kappa I^2)A^{2/3}$ and $E_C^0 = a_cZ^2A^{-1/3}$ are 
energies corresponding to spherical
shape. The relative surface and Coulomb energies
$B_s=E_s/E_s^0$, $B_C=E_C/E_C^0$ are only functions of the nuclear
shape. The dependence on the neutron and proton numbers is contained
in $E_s^0$ and in the fissility parameter $X=E_C^0/(2E_s^0)$. The constants
are $a_s=17.9439$~MeV, $\kappa =1.7826$, $a_c=3e^2/(5r_0)$, 
$e^2=1.44$~MeV$\cdot$fm, $r_0=1.2249$~fm.
\begin{figure}[ht]
\centerline{\epsfxsize=12cm\epsffile{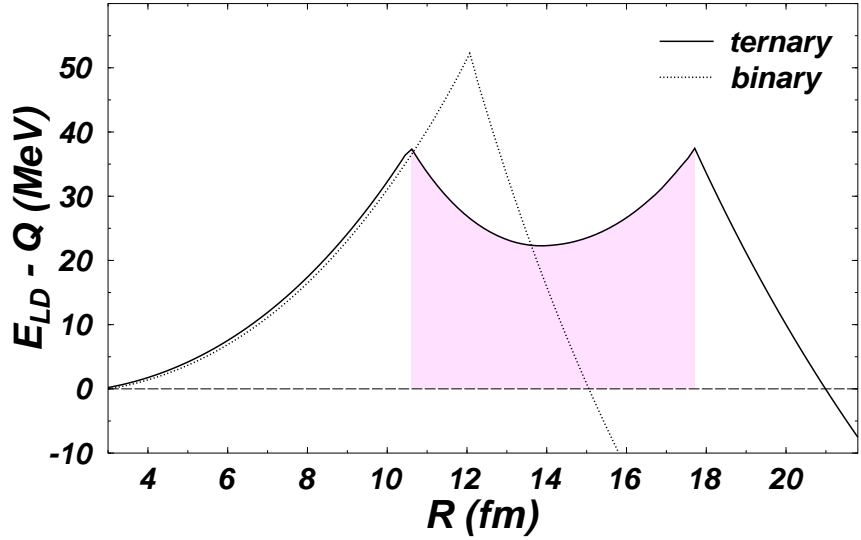}}
\caption{Deformation energy for the binary and ternary fission (accompanied by
$^{14}$C emission) of 
$^{240}$Pu. In both cases the heavy fragment is the double magic nucleus
$^{132}_{50}$Sn$_{82}$. The light fragments for binary- and ternary
processes are $^{108}$Ru and $^{94}$Sr, respectively. The region due to the
devolopment of the light particle,  from $R_{ov3}$ to $R_{t}$ is
emphasized.}
\label{Figure 3}
\end{figure}

To the deformation energy expressed in eq. (4), we add a small
phenomenological shell correction, 
allowing to reproduce, at $R=R_i$, exactly the experimental
$Q$-value in a system in which the origin of energy is taken as the sum of
self energies of the fragments separated at infinity. 
\begin{equation}
E_{LDM}(R)=E_{def}(R)+Q_{th}+(Q_{exp}-Q_{th})[1-(R-R_i)/(R_t-R_i)]-Q_{exp}
\end{equation}
which is $E_{def}(R)+(Q_{th}-Q_{exp})(R-R_i)/(R_t-R_i)$, where 
$Q_{th}=E^0-(E^0_1+E^0_2+E^0_3)=E^0_s+E^0_C-\sum_1^3(E^0_{si}+E^0_{Ci})$. 
In this manner the
barrier height increases if $Q_{exp}<Q_{th}$ and decreases if
$Q_{exp}>Q_{th}$. The correction is
increased gradually with $R$ up to $R_{t}$ and then remains constant for
$R>R_t$. Apart this correction, after the touching
point configuration, $R\geq R_{t}$, one is left with the Coulomb 
interaction energies. For spherical fragments this has the same expression
as that would be obtained for points placed into the fragment centers and
carying their whole charge.

Both the surface and Coulomb energies 
are calculated by performing numerical integration.\cite{dav75jcp,p75cpc78}
The relative surface energy is proportional to surface area. By
expressing the nuclear surface equation in cylindrical coordinates
$\rho = \rho (z, \varphi)$, one has  
\begin{equation}
B_s = \frac {1}{4\pi R_0^2}\int_{z'}^{z''}dz \int_{0}^{2\pi}
\rho \left [ 1 + \left (\frac {\partial \rho}{\partial z} 
\right ) ^2 + \left ( \frac {1}{\rho} \frac {\partial \rho}{\partial 
\varphi} \right ) ^2 \right ] ^{1/2}   d\varphi
\end{equation}
where $z', z''$ are the intersection points of the nuclear surface
with Oz axis.
Generally speaking, the Coulomb energy, $E_C$, for a system of three
fragments with different charge densities, is defined by the following
six fold integrals 
\begin{eqnarray}
E_C & = & \sum_1^3\frac { \rho _{ie}^2}{2} \int _{V_i} d^3r_1 \int _{V_i}
\frac {d^3r_2}{r_{12}}
+ \nonumber \\  &  &
\sum_{j \neq k} \rho _{je}\rho_{ke} \int _{Vj}d^3r_1 \int_{Vk}\frac 
{d^3r_2}{r_{12}} 
\end{eqnarray}
where the first three terms belong to individual fragments and the other 
three represent their interaction. Here $r_{12}=|{\bf r}_1 - {\bf
r}_2|$. The charge densities of the compound
nucleus and of the three fragments are denoted by $\rho _{0e}$,
$\rho _{1e}$, $\rho _{2e}$  and $\rho _{3e}$ respectively. The six-fold
integral is reduced to a four-fold one of the following kind. 
\begin{eqnarray}
E_C&=& \frac {\rho _e ^2}{10} \int _{z'} ^{z''} dz
\int _{z'} ^{z''} dz_1 \int _0 ^{2\pi} d\varphi \int _0 ^{2\pi} 
d\varphi _1 \left ( \rho ^2 - \frac {z}{2} \frac {\partial \rho ^2}
{\partial z} \right ) \left [\rho ^2 _1 - \right. \nonumber \\
 & &\left. \rho \rho _1 \cos (\varphi - \varphi _1) + \rho \frac {\partial
\rho _1}{\partial \varphi _1} \sin (\varphi - \varphi _1) +
\frac{(z-z_1)}{2} \frac{\partial \rho ^2 _1}{\partial z_1} \right ]
[\rho ^2 +  \nonumber \\
 & &\rho _1 ^2 - 2\rho \rho _1 \cos (\varphi - \varphi _1) +
(z-z_1)^2 ]^{-1/2}
\end{eqnarray}
for a general shape without axial symmetry.
One can get three-fold integrals for shapes possesing a 
symmetry axis, as for example:
\begin{equation}
B_{c1}= b_c \int _{-1} ^{x_c} dx \int _{-1} ^{x_c} dx' F(x, x') 
\end{equation}
where $b_c = 5d^5 /8\pi$, $d = (z'' - z')/2R_0$, and $x_c$ is the
position of separation plane between fragments with -1, +1 intercepts
on the symmetry axis (surface equation $y = y(x)$ or $y_1 = y(x')$).
In the integrand
\begin{eqnarray} 
F(x,x')&=&\{ y y_1[(K-2D)/3]\cdot \nonumber \\ & &
\nonumber \left [ 2(y^2+y_1^2)-(x-x')^2+\frac{3}{2}(x
-x')\left ( \frac{dy_1^2}{dx'}-\frac{dy^2}{dx} \right ) \right ]+
\nonumber \\
 & &K \left \{ y^2y_1^2/3+\left [y^2-\frac{x-x'}{2}\frac{dy^2}{dx}
\right ] \left [y_1^2-\frac{x-x'}{2}\frac{dy_1^2}{dx'}\right ] 
\right \} \} a_{\rho}^{-1} 
\end{eqnarray}  
$K$ and $K'$ are the complete elliptic integrals of the first and
second kind, respectively:
\begin{equation}
K(k) = \int _0^{\pi /2}(1-k^2 {\sin}^2 t)^{-1/2} dt 
\end{equation}
\begin{equation}
K'(k) = \int _0^{\pi /2}(1-k^2 {\sin}^2 t)^{1/2} dt 
\end{equation}
and $a_{\rho} ^2 = (y+y_1)^2+(x-x')^2$, $k^2 = 4yy_1 /a_{\rho}^2$, $D
= (K - K')/k^2$. In our computer program 
the elliptic integrals are calculated by using Chebyshev polynomial
approximation. For $x = x'$ the function $F$ is not determined. In
this case, after removing the indetermination, we get $F(x,x')=4y^3
/3$.

\section{Results}

Two examples of deformation energies are presented in Figures 2 and 3. They
were obtained for the $\alpha$-particle-(Fig. 2) and $^{14}$C (Fig. 3) 
accompanied fission of $^{240}$Pu, by assuming a double-magic heavy
fragment $^{132}_{50}$Sn$_{82}$. The corresponding deformation energy for
the binary cold fission of the same nucleus is also shown.

We would like to stress two striking features of these plots. Besides the
first (ground state) minimum there is a {\em second minimum}, proving the
nuclear molecule character of the aligned configuration of three fragments
in touch [($^{132}$Sn, $^{4}$He, $^{104}$Mo) and ($^{132}$Sn, $^{14}$C, 
$^{94}$Sr), respectively]. 

On the second hand, by comparing the surface areas under the deformation
energy curve of the binary and ternary pocesses, one can see the difference
explaining at least qualitatively the increased yield of the binary relative
to that of the ternary cold fission.

\end{document}